\documentclass[prd,nofootinbib,preprint,superscriptaddress]{revtex4-1}

\usepackage{amsmath, amssymb, amsthm, graphicx, epsfig, fancyhdr,epsfig}

\usepackage{tikzsymbols}
\usepackage{float}
\usepackage{xcolor}

\newcommand{\be}{\begin{equation}}
\newcommand{\ee}{\end{equation}}
\newcommand{\bea}{\begin{eqnarray}}
\newcommand{\eea}{\end{eqnarray}}

\newcommand{\ag}[1]{\textcolor{blue}{[Anish: #1]}}

\usepackage{tikz-feynman}
\tikzfeynmanset{compat=1.1.0}

\begin{document}

\title{ eV Hubble Scale Inflation with Radiative Plateau: \\ \it{Very light Inflaton,  Reheating \& Dark Matter in B-L Extensions} 
}

\author{Anish Ghoshal}
\email{anish.ghoshal@fuw.edu.pl}
\affiliation{Institute of Theoretical Physics, Faculty of Physics, University of Warsaw, ul. Pasteura 5, 02-093 Warsaw, Poland}

\author{Nobuchika Okada}
\email{okadan@ua.edu}
\affiliation{Department of  Physics  and  Astronomy, \\ University  of  Alabama,  Tuscaloosa,  Alabama  35487,  USA}

\author{Arnab Paul}
\email{arnabpaul9292@gmail.com  }
\affiliation{Indian Statistical Institute, \\ 203 B.T. Road, Kolkata-700108, India}

\begin{abstract}
\textit{ We study radiative plateau-like inflation \& Z$_{BL}$-portal freeze-in fermionic dark matter (DM) in a minimal B-L extended model. The U(1)$_{B-L}$ Higgs, responsible for heavy neutrino masses, also drives inflation in the early universe, thanks to radiative corrections from the heavy neutrinos \& the Z$_{BL}$ gauge boson. In our benchmark choice for the U(1)$_{B-L}$ gauge coupling $g_{B-L}\sim10^{-4}$, a light Z$_{BL}$ boson can be explored by current and future lifetime frontier experiments, such as FASER and FASER 2 at the LHC, SHiP, Belle II and LHCb. For the benchmark, the Hubble scale of inflation ($\mathcal{H}_{inf}$) is very low ($\mathcal{H}_{inf} = \mathcal{O}(100)$ eV) \& the inflaton turns out to be very light with mass of $\mathcal{O}(1)$ eV, and consequently the decay width of inflaton is extremely small. We investigate a 2-field system with the inflaton/B-L Higgs and the Standard Model (SM) Higgs, and find that the reheating with a suffuciently high temperature occurs when the water-fall direction to the SM Higgs direction opens up in the trajectory of the scalar field evolution.
}

\end{abstract}

\maketitle

\section{Introduction}
\label{intro}

It is well known since long that the candidate responsible for the initial accelerated expansion of the universe (inflation) may be associated with a scalar field (inflaton) \cite{Linde:1981mu, Guth:1980zm}. Although the exact nature of this field remains a mystery but with the plethora of cosmological data available we are in a position to study and distinguish between the ultra violet (UV) properties of various inflationary models. From particle theory \textit{point-of-view}, it is often fancied that the inflaton field
is also associated with solving other familiar issues in the Standard Model (SM). In this paper we investigate a model where the SM neutrinos receive mass in \textit{Type-I seesaw} framework, in a minimal gauged (B-L) extension, and provides a Z$_{BL}$-portal freeze-in dark matter (DM) candidate. The bosonic and fermionic quantum corrections, respectively, from the Z$_{BL}$ and the heavy neutrinos help to make the inflaton potential flat and inflation occurs in a radiative \textit{plateau-like} region. This gives us the very interesting possibility to verify inflationary predictions with laboratory searches in a UV-complete particle model. Moreover, this model contains a promising dark matter candidate, the so-called $Z_{BL}-$portal fermion DM.

Inflation scenarios driven by the quartic potential of inflaton is highly disfavored by the CMB data as they predict too large tensor to scalar ratio \cite{Akrami:2018odb} \footnote{By introducing a coupling of the inflaton field to Ricci scalar (non-minimal gravitational coupling), tensor to scalar ratio $r$ values can be lowered \cite{Bezrukov:2007ep, Libanov:1998wg, Fakir:1990eg, Futamase:1987ua, Masina:2018ejw,Okada:2010jf, Linde:1981mu, Kallosh:1995hi, Inagaki:2014wva,Ghoshal:2022qxk}.}. The so-called inflection-point inflation is a diligent way to fit the CMB power-spectrum by flattening the effective inflaton quartic potential with radiative corrections from bosonic and fermionic loops (among other ways) \cite{Ballesteros:2015noa, NeferSenoguz:2008nn, Enqvist:2013eua, Okada:2017cvy, Choi:2016eif,Allahverdi:2006iq,Allahverdi:2006we,BuenoSanchez:2006rze,Baumann:2007np,Baumann:2007ah,Badziak:2008gv,Enqvist:2010vd,Cerezo:2012ub,Choudhury:2013jya,Choudhury:2014kma,Ballesteros:2015noa,Choi:2016eif,Okada:2016ssd,Okada:2017cvy,Stewart:1996ey,Stewart:1997wg,2104.03977,Ghoshal:2022jeo}\footnote{ See Refs. \cite{Biondini:2020xcj,SravanKumar:2018tgk} for inflation in GUT framework with DM.}. It was shown that even with a negligible non-minimal coupling values the model can fit the CMB data\footnote{Of course a non-minimal gravitational coupling should be nevertheless be present in a complete analysis, since it is radiatively generated, even if it is set to zero at some scale. Nevertheless, such a coupling is only important in the trans-Planckian regime.}.

Using bosonic and fermionic corrections to achieve the inflection-point was studied in Refs. \cite{Okada:2017cvy, Choi:2016eif, Enqvist:2010vd};  particularly, B-L gauged U(1)-Higgs inflation has been studied in this set-up, giving the predictions consistent with the CMB observations \cite{Okada:2016ssd,Okada:2019yne,Ghoshal:2022hyc} as well as a collider search for a long-lived inflaton \cite{Okada:2019opp}. As we will see 
inflation demands a tiny gauge coupling which in turn dictates us to embed the inflationary set-up in a small field inflation scenario, resulting a small inflaton mass and incurring the following:
\begin{enumerate}
    \item If the reheating proceeds via the inflaton decay, the inflaton is too light to have produced the visible universe.
    \item 
    The tiny gauge coupling makes it hard for the Z$_{BL}$-portal DM particle to get in the thermal plasma of the SM particles.
\end{enumerate}

In this article we study the inflection-point inflation and how to resolve the aforementioned issues by closely scrutinizing the early universe dynamics. We will show that even if the inflaton is extremely light it is possible to have a reheating temperature to be sufficiently high due to opening up of the waterfall direction in the SM Higgs direction in the scalar field trajectory of oscillation after inflation. 
 
 For the parameter choice in this work, we find that for a typical DM mass of $m_\chi=\mathcal{O}(10)$ GeV, $Z_{BL}$ is thermalised in the SM plasma at $T\sim m_\chi$. By setting a suitable $B-L$ charge for DM, DM ($\chi$) can be produced with the right abundance via the freeze-in mechanism. We discuss a falsifiability of the model in future experiments. Due to the smallness of the coupling strengths, it is very challenging to see any direct laboratory signatures of such a freeze-in DM. However, one of the plausible pathways to detect the freeze-in DM is to have the DM production in the early Universe via the decays of dark sector particles which are in thermal equilibrium with the SM plasma. Such feeble couplings associated with the decays make the dark sector particles very long-lived and can be looked for at the Large Hadron Collider (LHC) and beyond (see~\cite{Curtin:2018mvb,Hessler:2016kwm,Belanger:2018sti}). On the other hand, if the freeze-in DMs are produced via scattering processes with a dark sector particle as a mediator, then depending on the nature of the portal (scalar-portal, vector-portal, etc.) one can search for the mediator particles at the lifetime frontier and intensity frontier experiments  \cite{Hambye:2018dpi,Heeba:2019jho,Mohapatra:2019ysk,Mohapatra:2020bze,Barman:2021lot,Barman:2021yaz,Das:2019fee,Chiang:2019ajm,Das:2018tbd,Ghoshal:2022ruy,Barman:2022scg,Barman:2022njh}\footnote{See Ref. \cite{Elor:2021swj} for other tests of freeze-in DM.}.

The paper is arranged as follows: in the next section, we describe a simple $B-L$ gauged extension of the Standard Model (SM), and describe how a plateau region can be created in the $B-L$ Higgs potential due to quantum corrections from the heavy neutrinos and the $B-L$ gauge boson. In the following section, we investigate the scalar field trajectory after inflation to see how a successful reheating can occur in a \textit{2-field} system with the $B-L$ and SM Higgs fields. Next, we discuss the formation of dark matter (DM) via the freeze-in mechanism and experimental searches in the model parameter space and end our study with conclusions in the final section.

\medskip

\section{Creating Point of Inflection in Gauged Higgs Potential}


In order to realize the inflection-point in a Higgs potential, we employ the Renormalization-Group (RG) improved effective Higgs potential. For a quartic potential, as the quartic coupling first decreases (due to fermionic corrections) and then increases in the UV (due to bosonic corrections). It is well-known that in the vicinity of the minimum point of the running quartic coupling, both the quartic coupling and its beta-function become vanishingly small due to which an inflection point can be realized in the effective Higgs potential. It is at this point that the relation between high energy and low energy physics become manifest: the quantum corrections in a particle model determine the point of inflection and the scale of inflation.

\subsection{The Model}

Minimal $B-L$ extension of the SM provides the ingredients for neutrino masses and lepton number violation \cite{Mohapatra:1980qe,Marshak:1979fm,Wetterich:1981bx,Masiero:1982fi,Mohapatra:1982xz,Davidson:1978pm}. Following \cite{Mohapatra:2019ysk}, we add a dark Dirac fermion $\chi$ (SM singlet, but charged under $U(1)_{B-L}$) as our DM candidate via the $B-L$ gauge interaction ($Z_{BL}$-portal).

\begin{table}[H]
\begin{center}
\begin{tabular}{c|ccc|c}
            & SU(3)$_c$ & SU(2)$_L$ & U(1)$_Y$ & U(1)$_{B-L}$  \\
\hline
$ q_L^i $    & {\bf 3}   & {\bf 2}& $+1/6$ & $+1/3$  \\ 
$ u_R^i $    & {\bf 3} & {\bf 1}& $+2/3$ & $+1/3$  \\ 
$ d_R^i $    & {\bf 3} & {\bf 1}& $-1/3$ & $+1/3$  \\ 
\hline
$ \ell^i_L$    & {\bf 1} & {\bf 2}& $-1/2$ & $-1$  \\ 
$ N_R^i$   & {\bf 1} & {\bf 1}& $ 0$   & $-1$  \\ 
$ e_R^i  $   & {\bf 1} & {\bf 1}& $-1$   & $-1$  \\ 
\hline 
$ H$         & {\bf 1} & {\bf 2}& $-1/2$  &  $ 0$  \\ 
$ \varphi$      & {\bf 1} & {\bf 1}& $  0$  &  $+2$  \\
\hline
$ \chi$      & {\bf 1} & {\bf 1}& $  0$  &  $Q_\chi$
\end{tabular}
\end{center}
\caption{ \it
Particle content of the model. $i=1,~2,~3$ is the generation index.
}
\end{table}

The right-handed neutrinos ($N_R^i$) have Majorana Yukawa interaction terms, 
\bea  
   {\cal L} \supset  - \frac{1}{2} \sum_{i=1}^{3} Y_i  \varphi  \overline{N_{R}^{i~C}} N^i_{R}  +{\rm h.c.}
\eea  
Associated with the gauge symmetry breaking, all the new particles, $B-L$ gauge boson ($Z_{BL}$), the right-handed neutrinos ($N_R$) and the $B-L$ Higgs acquire their masses, which are as follows:
\bea 
m_{Z_{BL}}= 2 \; g_{BL} \; v_{BL},  \; \; 
m_{N^i}= \frac{1}{\sqrt{2}} \; Y_i \; v_{BL}, \; \; 
m_{\phi} = \sqrt{2 \lambda_\phi} \;  v_{BL}, 
\label{masses}
\eea
where $\lambda_\phi$ is the quartic coupling of the $B-L$ Higgs field $\varphi$ (see the next section), and $v_{BL}= \sqrt{2} \langle \varphi\rangle$ is the VEV of the $B-L$ Higgs field. The $B-L$ Higgs field $\varphi$ can be redefined as $\varphi=(\phi+v_{BL})/\sqrt{2}$ in the unitary gauge, and we identify the real scalar field $\phi$ to be the inflaton.

Although when the SM is extended with the gauged $U(1)_{B-L}$ symmetry, there is no anomaly associated with the $B-L$ gauge symmetry due to the presence of three SM-singlet right-handed neutrinos with $B-L$ charges of $-1$. On top of this, as mentioned before, we introduce an SM singlet Dirac fermion $\chi$: our DM candidate. 
Interaction part of the Lagrangian includes, 
\begin{eqnarray}
{\cal L}_{Z_{BL}} &=& y_l\,\bar{L}\Tilde{H} N  + g_{BL}\,\left( Z_{BL} \right)_\mu \left[\sum_f (B-L)_f \bar{f}\gamma^\mu f
+Q_\chi \bar{\chi}\gamma^\mu \chi~\right]~+~{\rm h.c}.
\end{eqnarray} 
where  $g_{BL}$, $Q_\chi$, $m_{Z_{BL}}$, and $m_\chi$ are the free parameters of the model. 



\subsection{Slow-Roll Parameters and Constraints from Planck 2018}

The inflationary slow-roll parameters are given by,
\begin{eqnarray}
\epsilon(\phi)=\frac{ M_{P}^2}{2} \left(\frac{V'}{V}\right)^2, \; \; 
\eta(\phi)=
M_{P}^2\left(\frac{V''}{V }\right), \;\;
\varsigma^2{(\phi)} = M_{P}^4  \frac{V^{\prime}V^{\prime\prime\prime}}{V^2}, \label{SRCond}
\end{eqnarray}
where $M_{P}= M_{Pl}/\sqrt{8 \pi} = 2.43\times 10^{18}$ GeV is the reduced Planck mass, $V$ is the inflation potential, and the prime is derivative with respect to inflaton $\phi$.  

The curvature perturbation $\Delta^2_{\mathcal{R}}$ is given by
\begin{equation} 
\Delta_{\mathcal{R}}^2 = \frac{1}{24 \pi^2}\frac{1}{M_P^4}\left. \frac{V}{ \epsilon } \right|_{k_0},
 \label{PSpec}
\end{equation}
 which must satisfy $\Delta_\mathcal{R}^2= 2.189 \times10^{-9}$
  from the Planck 2018 results \cite{Akrami:2018odb} at pivot scale $k_0 = 0.05$ Mpc$^{-1}$.
The number of e-folds is given by
\begin{eqnarray}
N=\frac{1}{M_{P}^2}\int_{\phi_E}^{\phi_I}\frac{V }{V^\prime} d\phi  ,
 \label{EFold}
\end{eqnarray} 
where $\phi_I$ is the value of inflaton during horizon exit of the scale $k_0$, 
  and $\phi_E$ is defined as the value of inflaton when slow-roll condition is violated, 
  i.e. $\epsilon(\phi_E)=1$.
  
The slow-roll approximation holds when
   $\epsilon \ll 1$, $|\eta| \ll 1$ , and $\varsigma^2\ll1$. 
The inflationary predictions are given by
\bea
n_s = 1-6\epsilon+2\eta, \; \; 
r = 16 \epsilon , \;\;
\alpha = 16 \epsilon \eta -24 \epsilon^2-2 \varsigma^2, 
 \label{IPred}
\eea 
where $n_{s}$ and $r$ and $\alpha \equiv \frac{\mathrm{d}n_s}{d ln k}$ are the scalar spectral index, the tensor-to-scalar ratio and the running of the spectral index, respectively, at $\phi = \phi_I$.  
The Planck 2018 results \cite{Akrami:2018odb} gives an upper bound on  $r \lesssim 0.067$,  
the bounds for the spectral index ($n_s$) and the running of spectral index ($\alpha$) are $0.9691 \pm 0.0041$ and $0.0023\pm 0.0063$, respectively. A joint study with dataset of Planck, BICEP/Keck 2018 and Baryon Acoustic Oscillations pulls down the upper bound of tensor-to-scalar ratio to $r < 0.032$ \cite{Tristram:2021tvh}. Future precision measurements have the potential to pin down the error in $\alpha$ to $\pm0.002$ \cite{Ade:2018sbj,Aiola:2020azj}.


\subsection{Achieving the Plateau}


The scalar potential for inflection-point inflation at an inflection point near $\phi = M$ is given by \cite{Okada:2016ssd}
\bea \label{pot}
V (\phi)\simeq V_0 +\sum_{n = 1}^3 \frac{1}{n!}V_n (\phi-M)^n , 
\label{eq:PExp}
\eea
   where $V_0 = V(M)$ is a constant, $V_n \equiv  d^{n}V/ d \phi^n |_{\phi =M}$ are derivatives evaluated at $\phi=M$, and $\phi = M$ is the inflaton field value at the pivot scale $k_0= 0.05$ Mpc$^{-1}$ of the Planck 2018 measurements \cite{Akrami:2018odb}. If the values of $V_1$ and $V_2$
   are tiny enough, the (almost) inflection-point can be realized. Re-writing Eqs.~(\ref{SRCond}) and (\ref{eq:PExp}) we get 
\bea
\epsilon(M) \simeq \frac{M_{P}^2}{2}\Big(\frac{V_1}{V_0}\Big)^2, \;\;
\eta(M) \simeq M_{P}^2\Big(\frac{V_2}{V_0}\Big), \;\;
\chi^2{(M)} = M_{P}^4  \frac{V_1 V_3}{V_0^2}, 
\label{IPa}
\eea
where we have used the approximation $V(M)\simeq V_0$. 
Similarly, the power-spectrum $\Delta_{\mathcal{R}}^2$ is expressed as
\bea
\Delta_{\mathcal{R}}^2 \simeq \frac{1}{12\pi^2}\frac{1}{M_P^6}\frac{V_0^3}{V_1^2}.
\label{CV1} 
\eea 
Using the observational constraint, $\Delta_{\mathcal{R}}^2= 2.189 \times  10^{-9}$, and a fixed $n_s$ value, we obtain 
\bea
\frac{V_1}{M^3}&\simeq& 1963\left(\frac{M}{M_P}\right)^3\left(\frac{V_0}{M^4}\right)^{3/2}, \nonumber \\
\frac{V_2}{M^2}&\simeq& -1.545\times 10^{-2}\Big(\frac{1-n_s}{1-0.9691}\Big)\Big(\frac{M}{M_P}\Big)^2\left(\frac{V_0}{M^4}\right), 
\label{FEq-V1V2}
\eea
where we have used $V(M)\simeq V_0$ and $\epsilon(M) \ll \eta(M)$. 
For the remainder of the analysis we set $n_{s}=0.9691$, the central value from the Planck 2018 results \cite{Akrami:2018odb}. We neglect the $O(10\%)$ level uncertainty in the values of cosmological inflationary parameters in this analysis as it will not change the highlight of this work. 
Then V$_3$ becomes
\bea
\frac{V_3}{M} \simeq 6.983 \times 10^{-7}\Big(\frac{60}{N}\Big)^2 \Big(\frac{V_0^{1/2}}{M M_P}\Big) . 
\label{FEq-V3} 
\eea

\smallskip

Using Eqs.~(\ref{IPred}), (\ref{FEq-V1V2}), (\ref{FEq-V3}), 
(\ref{IPa}) and (\ref{FEq-V1V2}), tensor-to-scalar ratio ($r$) is given by
\bea 
r=3.082\times 10^7\Big( \frac{V_0}{M_P^4}\Big). 
\label{FEq-r}
\eea 
and, the running of the spectral index ($\alpha$) 
\bea
\alpha \simeq - 2\varsigma^2(M) = - \; 2.741 \times 10^{-3}\left(\frac{60}{N}\right)^2, 
\label{FEq-alpha}
\eea
Note that the running is independent of $V_0$ and $M$. This prediction is consistent with the  current experimental bound, $\alpha=0.0023\pm 0.0063$ \cite{Akrami:2018odb}. 
Precision measurement of the running of the spectral index in future experiments can reduce the error to $\pm0.002$ \cite{Ade:2018sbj,Aiola:2020azj}. 
Hence, the prediction can be tested in the future. 


\subsection{Radiative Plateau}

For creating the plateau in the U(1)$_{B-L}$ Higgs potential $V_{tree}= \frac{1}{4}  \lambda_{\phi-tree} \phi^4$, let us take a look at the RGE improved effective potential:  
\bea
V(\phi) = \frac{1}{4} \lambda_\phi (\phi)\;\phi^4, 
\label{VEff}
\eea
where $\lambda_\phi (\phi)$ is the solution to the RGEs which involves the beta functions of $g_{BL}$, $Y$ and $\lambda_\phi$, $\beta_{g_{BL}}$, $\beta_{Y}$ and $\beta_{\lambda_\phi}$, respectively.
For simplicity, we assume the degenerate mass spectrum for the right-handed neutrinos, $Y \equiv Y_1 = Y_2=Y_3$. The coefficients in the expansion of Eq.~(\ref{eq:PExp}) is given as \footnote{For details of this derivation, see Ref. \cite{Okada:2016ssd}.}: 
\bea
\frac{V_1}{M^3}&=& \frac{1}{4} (4 \lambda_\phi + \beta_{\lambda_\phi}),\nonumber \\
\frac{V_2}{M^2}&=& \frac{1}{4} (12\lambda_\phi + 7\beta_{\lambda_\phi}+M \beta_{\lambda_\phi}^\prime), \nonumber \\
\frac{V_3}{M}&=& \frac{1}{4} (24\lambda_\phi + 26\beta_{\lambda_\phi}+10M \beta_{\lambda_\phi}^\prime+M^2 \beta_{\lambda_\phi}^{\prime\prime}), 
\label{ICons2}
\eea
where the prime denotes the differential coeffcient $d/d\phi$.
Using $V_1/M^3\simeq 0$ and $V_2/M^2\simeq 0$, we obtain
\bea
 \beta_{\lambda_\phi} (M)\simeq -4\lambda_\phi(M), \qquad
 M\beta_{\lambda_\phi}^{\prime}(M)\simeq 16 \lambda_\phi (M). 
 \label{Cond1}
\eea
Hence the last equation in Eq.~(\ref{ICons2}) is simplified to $V_3/M \simeq 16 \;\lambda_\phi(M)$. 
Comparing it with Eq.~(\ref{FEq-V3}), we obtain 
\bea
\lambda_\phi(M)\simeq 4.762 \times 10^{-16} \Big(\frac{M}{M_{P}}\Big)^2\Big(\frac{60}{N}\Big)^4,
\label{FEq1} 
\eea 
where we have approximated $V_0\simeq (1/4) \lambda_\phi(M) M^4$. 
Since the $\lambda_\phi(M)$ is extremely small, we approximate $\beta_{\lambda_\phi}(M) \simeq 0$, which leads to 
\bea
Y(M)\simeq {32}^{1/4}\;g_{BL}(M). 
\label{FEq3}
\eea
For this relation between $g_{BL}$ and $Y$, we have assumed that the gauge and the Yukawa couplings 
dominate the beta-function, as a consequence the mass ratio of the right-handed neutrinos and the $B-L$ gauge boson is fixed, in order to have a plateau inflation. 

\begin{figure}[H]
\begin{center}
\includegraphics[height=0.4\textwidth]{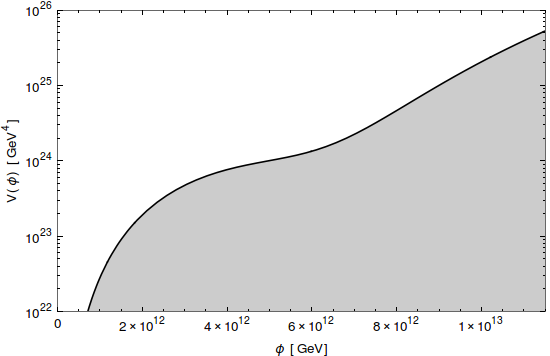}
	\caption{\it 
	 RG improved $\rm B - L$ inflaton potential as a function of $\phi$. 
	 Here, we have fixed $M= 2.78\times10^{-6}~ M_P$, so that $g_{BL}=10^{-4} ,~Y= 2.3\times10^{-4}~ \& ~ \lambda_{\phi} (M) =3.7\times10^{-27} $. We note that the inflection-point-like point appears at $\phi \sim M$.}
\label{fig:gwbc}
\end{center}
\end{figure}

Using the second equation in Eq.~(\ref{Cond1}) and Eq.~(\ref{FEq3}), we find $\lambda_\phi(M)\simeq 3.713\times 10^{-3} \;g_{BL}(M)^6$. 
Then from Eq.~(\ref{FEq1}), $g_{BL}(M)$ is expressed as 
\bea
g_{BL}(M)\simeq 7.107\times 10^{-3} \;\Big(\frac{M}{M_{P}}\Big)^{1/3}.
\label{FEq2} 
\eea
Finally, from Eqs.~(\ref{FEq-r}) and (\ref{FEq1}), the tensor-to-scalar ratio ($r$) is given by 
\bea
r \simeq 3.670 \times 10^{-9}  \Big(\frac{M}{M_{P}}\Big)^6, 
\label{FEqR} 
\eea
which is very small, as expected in the single field inflationary scenarios. 
For a sample parameter choice, we plot the effective potential in Fig. \ref{fig:gwbc}.


\medskip

\section{Reheating Dynamics of the Scalar Fields}
\label{sec:reheat}



\begin{figure}
\includegraphics[width=0.31\textwidth]{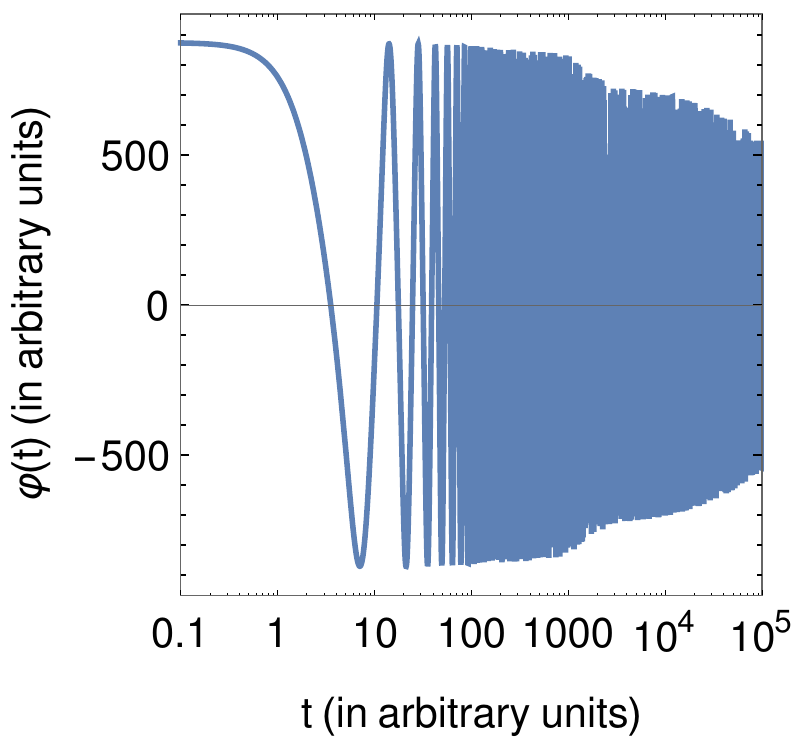}
\includegraphics[width=0.3\textwidth]{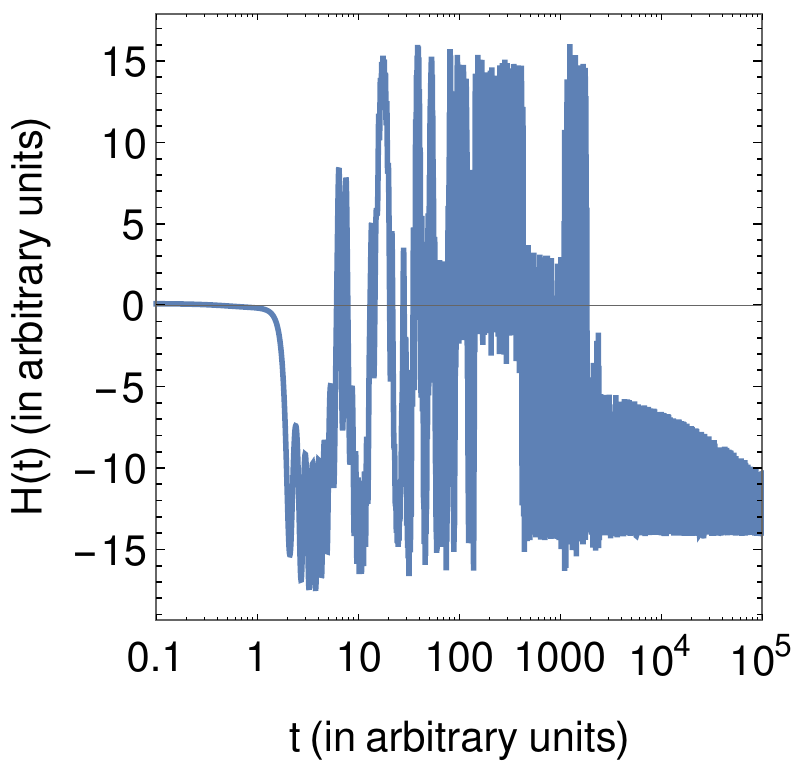}
\includegraphics[width=0.32\textwidth]{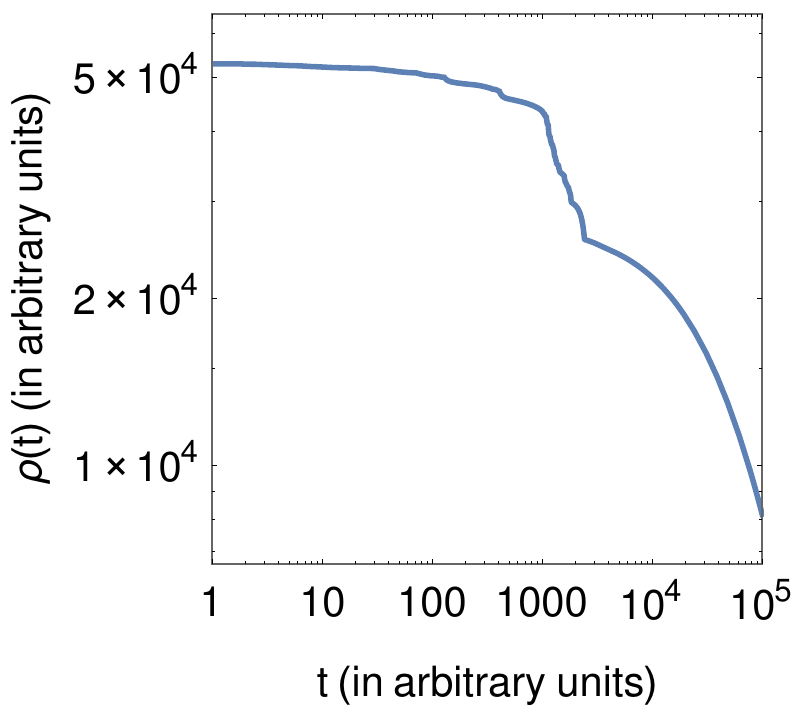}
\caption{\it  Evolution of the two fields $\varphi (t)$ and $H (t)$  and the energy density $\rho (t)$ with time $t$ for a toy model.}
\label{fig:phi and h}
\end{figure}

\begin{figure}
\begin{center}
\includegraphics[width=0.5\textwidth]{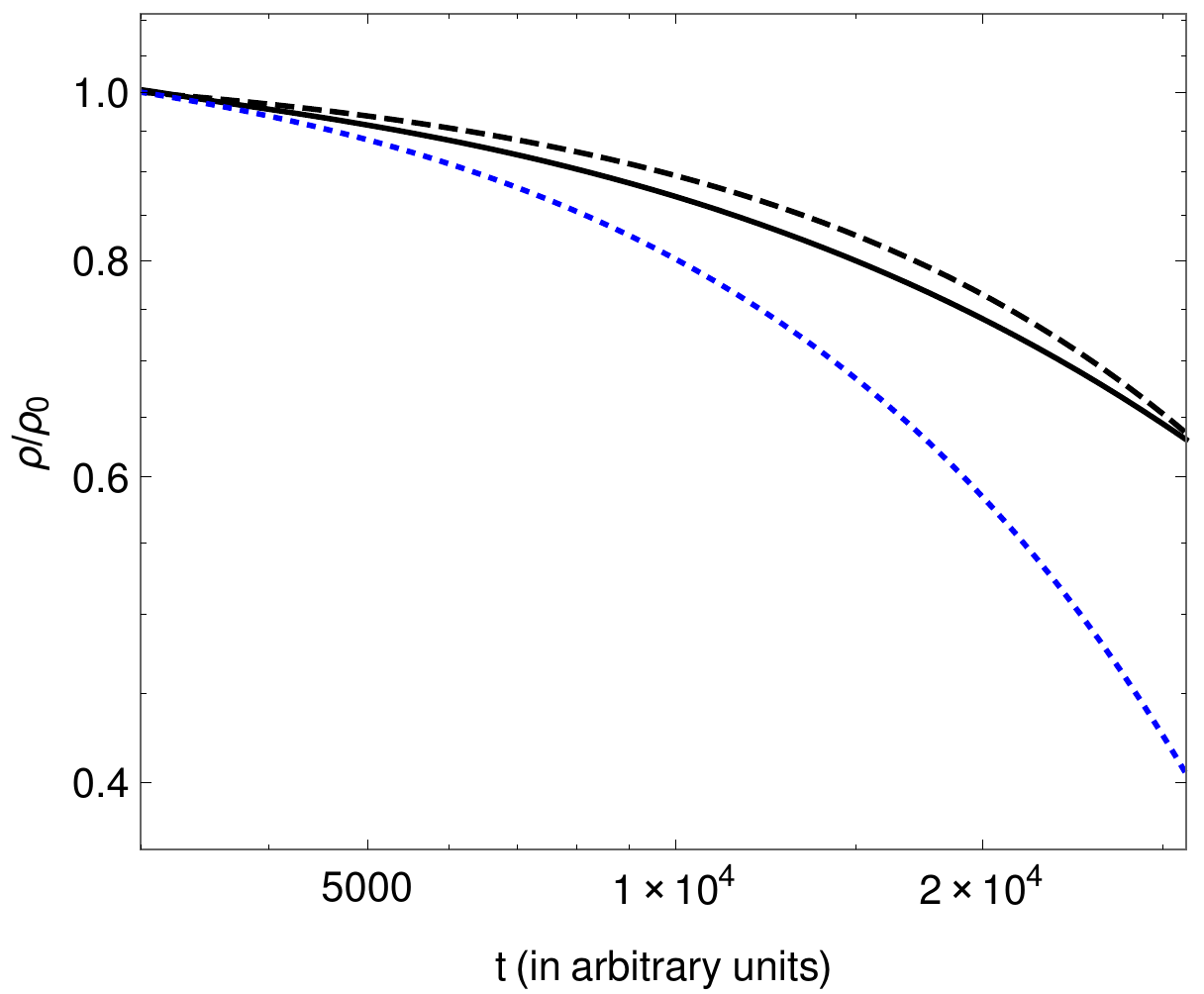}
\caption{\it Black solid curve denotes the numerical solution of $\rho$ vs t for the toy model (with a normalisation factor $\rho_0$). The Blue dotted curve is $e^{-\Tilde{\Gamma}_{eff}t}$, whereas the Black dashed curve is $e^{-0.5~\Tilde{\Gamma}_{eff}t}$. It is clear from these plots that the approximate formula of $\Gamma_{eff}$ given by Eq. (\ref{Gammaeff}) describes the decay within a factor of 2.
}
\label{fig:rho compare}
\end{center}
\end{figure}

In order to investigate the reheating dynamics, we start with the two-dimensional field space with scalar potential of the model given by
\bea
\label{InfPot}
V(|H|,|\varphi|) =  \lambda_\phi \left(|\varphi|^2 -\frac{v^2_{BL}}{2}\right)^2 + \lambda_H \left(|H|^2 -\frac{v^2_H}{2}\right)^2 \\ \nonumber +  \sqrt{4 \lambda_{H} \lambda_{\phi}  \xi} \left(|H|^2 -\frac{v^2_H}{2}\right)\left(|\varphi|^2 -\frac{v^2_{BL}}{2}\right),
\eea 
where $H$ is the SM Higgs, $\varphi$ is the inflaton ($U(1)_{B-L}$-gauged Higgs), $v_{H}$ \& $v_{BL}$ are their VEVs, respectively, $\lambda_\phi$ and $\lambda_H$ are their respective quartic couplings and $\sqrt{4 \lambda_{H} \lambda_{\phi}  \xi}$ is the quartic-mixing of $\varphi$ and $H$. 
\begin{figure}
\begin{center}
\includegraphics[width=0.5\textwidth]{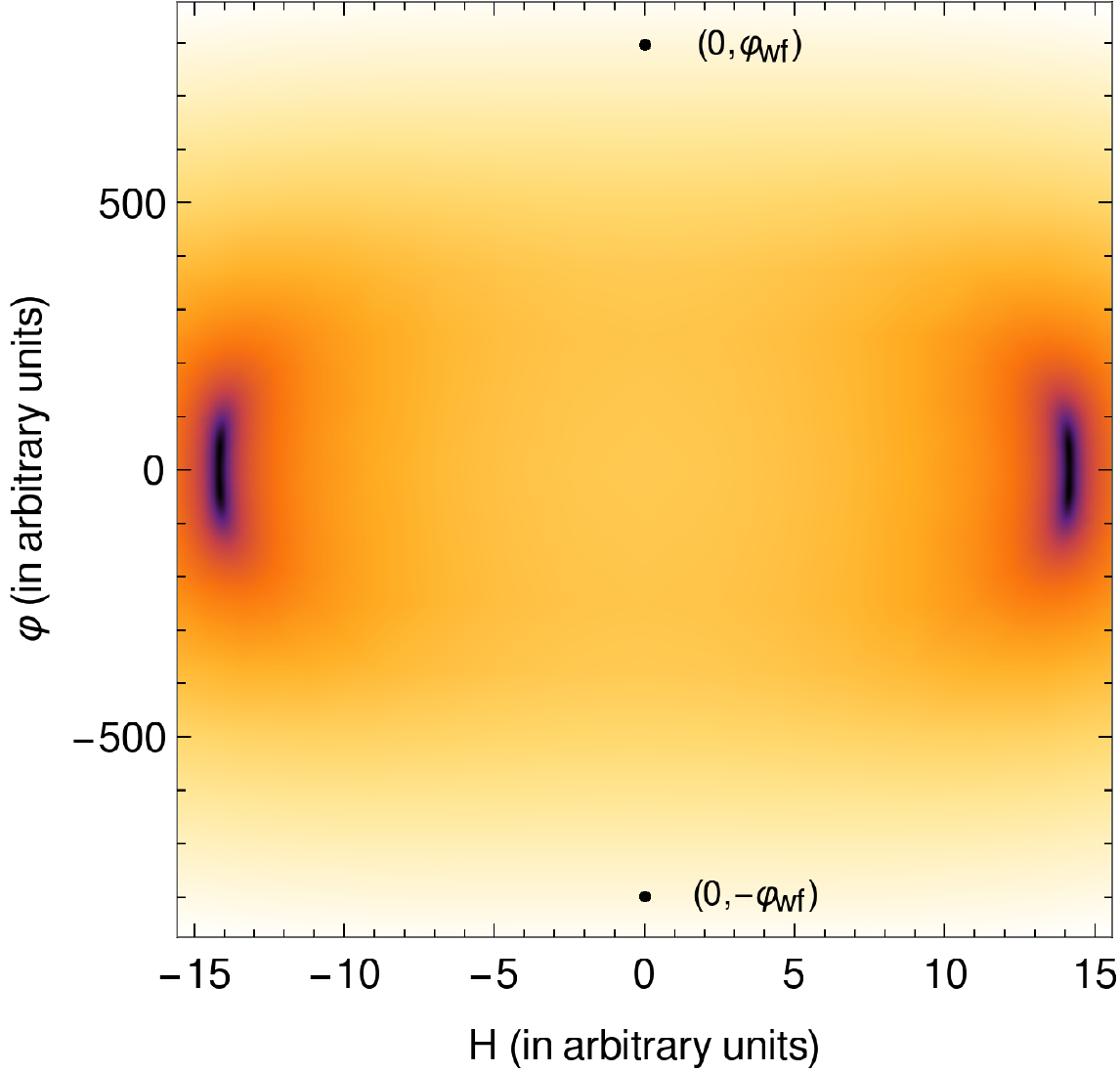}
\includegraphics[width=0.049\textwidth]{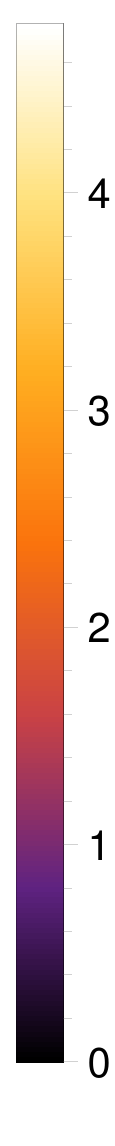}
\caption{\it  Plot of  $\log_{10}(V+1)$ as function of $H$ and $\varphi$ (toy model) in color coding. 
}
\label{fig:potdensity}
\end{center}
\end{figure}
After inflation the field rolls down along the inflaton direction towards $\varphi=0$. We numerically solve the equations of the motion of this 2-field system with the SM Higgs boson decay width $\Gamma_H$, but neglecting the inflaton decay width. The equations are given by
\begin{eqnarray}\label{eom}
  \ddot\varphi+3\mathcal{H}\dot{\varphi}+V_{,\varphi}&=&0,\\
    \ddot{H}+(3\mathcal{H}+\Gamma_H)\dot{H}+V_{,H}&=&0.
\end{eqnarray}
Although there is a ``bump" (local maximum) 
 around (0,0) in the potential as shown in Fig.~\ref{fig:potdensity}, at the initial stage, due to high inertia in the $\varphi$ direction, the oscillation occurs mostly along the $\varphi$ direction. With time, this oscillation amplitude along $\varphi$ dies down slowly due to Hubble friction until the $\varphi$ amplitude becomes similar to the order of the length of the potential bump along $\varphi$ axis. At this stage, the field feels the slope of the potential around the potential bump. During this oscillation, before the field reaches the $\varphi=0$ value, the waterfall direction opens up in the SM Higgs direction, which is at
\be
\varphi=\varphi_{wf}\equiv \frac{1}{\sqrt{2}}\sqrt{v_{BL}^2+\frac{\lambda_H v_H^2}{\sqrt{\lambda_H \lambda_\phi  \xi }}}.
\ee
If the field is displaced by a tiny amount from the inflaton ($\varphi$) axis (due to fluctuations), the field rolls down the waterfall direction (with additional rapid oscillation parallel to the SM Higgs direction) and the field follows an approximately semi-elliptic path (this path is truly semi-elliptic for $\xi=1$) of semi-major axis $2\times\varphi_{wf}$ and semi-minor axis $2\times H_{minor}$, where
\be
H_{minor}\equiv \frac{\sqrt{v_{BL}^2 \sqrt{\lambda_H \lambda_\phi  \xi }+\lambda_H v_H^2}}{ \sqrt{2\lambda_H}},
\ee
around the (0,0) point of the field space to reach $\varphi=-\varphi_{wf}$.
The additional oscillation in the SM Higgs direction about the smooth semi-elliptical path dies down quickly due to decay rate of SM Higgs $\Gamma_H$, after which only the smooth elliptical path dynamics is left. The trajectory of the field from the numerical solutions is shown in Fig. \ref{fig:phi and h} with a set of arbitrary values of parameters (termed as ``toy model" 
in Figs.~\ref{fig:phi and h} and \ref{fig:rho compare}).
The ``toy model" is chosen in such a way that the total time scale ($\Delta t_{sol}$) we solved the equations (\ref{eom}) for, satisfies $\mathcal{H}_{wf}^{-1}\gg \Delta t_{sol}\sim\Tilde{\Gamma}_{eff}^{-1}$, where $\mathcal{H}_{wf}$ and $\Tilde{\Gamma}_{eff}$ are defined in Eq.~(\ref{eq:Trh}) and Eq.~(\ref{Gammaeff}), respectively. This choice enables us to observe the effect of energy dilution in the fields resulting directly from the decay, not due to Hubble friction. This choice ($\mathcal{H}_{wf}^{-1}> \Tilde{\Gamma}_{eff}^{-1}$) is also satisfied in the realistic case we discuss later on.
Although the values of the parameters used for the plots in this subsection are arbitrary (toy model), the physical characteristics of the dynamics are similar for our realistic choice of the parameters.

As the tangent vector of this approximate semi-elliptic path in the \textit{2-field system} has a component in the SM Higgs direction, even if the decay rate of inflaton $\Gamma_\phi=0$ \footnote{Note that the mass of the inflaton $m_\phi=\sqrt{2\lambda_\phi}v_{BL}$ is very small for small values of $\lambda_\phi$, possibly making it kinematically inaccessible for the inflaton to decay in most channels through mixing with SM Higgs (with mixing angle $\theta\sim\sqrt{\xi}\frac{m_\phi}{m_H}$).}, the oscillation along the elliptical path dies down quickly due to the ``effective decay rate" arising from the friction in the SM Higgs direction:
\be
\Gamma_{eff}\sim \Gamma_H \left(\frac{H_{minor}}{\varphi_{wf}}\right)^2\equiv \Tilde{\Gamma}_{eff}.
\label{Gammaeff}
\ee
This quick depletion of energy density starting from near $t=2000$ (in arbitrary units), as shown in Fig. \ref{fig:phi and h}, can be interpreted as the reheating of the universe.

We see in Fig. \ref{fig:rho compare} that this approximate formula of $\Gamma_{eff}$ mimics the numerically estimated energy depletion within a factor of $\mathcal{O}(1)$. If Hubble parameter at the water-fall $\mathcal{H}_{wf}\ll \Gamma_{eff}$, i.e. decay time-scale is negligible with respect to Hubble time-scale, we can assume instantaneous reheating. We then estimate the reheating temperature as
\be
T_{rh}=0.55\left(\frac{100}{g_*}\right)^{1/4}\sqrt{\mathcal{H}_{wf}~M_P}~\left({\rm{when~}}\Gamma_{eff}\gg\mathcal{H}_{wf}\equiv \frac{1}{\sqrt{3}M_P}\sqrt{V(0,\varphi_{wf}) }\right).
\label{eq:Trh}
\ee
We emphasise that, the non-trivial field dynamics and hence the effective decay rate $\Gamma_{eff}$ enables the reheating temperature to be greater than the BBN energy scale, even if the actual decay rate of the inflaton itself is unable to do so. Below we give a benchmark point for our study.

\subsection{Case Study for Benchmark Point}


For our benchmark point $g_{BL}= 10^{-4},~Y=2.3\times10^{-4},~\lambda_\phi=3.7\times10^{-27}$ during inflation, at lower energy\footnote{evolving via the RGE.} 
for $\lambda_\phi=1.8\times10^{-23},~v_{BL}=10^3$ GeV, $\lambda_H=0.1,~v_H=246$ GeV, $\xi=0.1$, we get $\Gamma_{eff}=1.72\times10^{-14}$ GeV. 
We note that, for our choice of the bench-mark point, the mass of inflaton is $m_\phi=\sqrt{2\lambda_\phi}v_{BL}\sim6\times10^{-9}$ GeV and inflation happens at Hubble parameter $\mathcal{H}_{inf}\sim7\times10^{-7}$ GeV (scale of inflation). For this value of $m_\phi$, inflaton may decay only into photons or neutrinos (via mixing with the SM Higgs). 
The mixing angle between inflaton and SM Higgs is also negligible at this value: $\theta\sim\sqrt{\xi}\frac{m_\phi}{m_H}\sim1.5\times10^{-11}$.
Hence the inflaton decay rate is very suppressed and as a result reheating of the universe from straight-forward decay of inflaton is difficult, and so is its compatibility with the BBN constraints. 

However, as we showed via numerical estimates in the toy model in section \ref{sec:reheat}, when the waterfall direction opens up, the Hubble parameter becomes $\mathcal{H}_{wf}=6.9\times 10^{-15}$ GeV, which is smaller than $\Gamma_{eff}$. This means that during this period, within one Hubble time, the energy density stored in the fields dilutes away, and the reheating temperature is approximated to be, $T_{rh}=0.55\left(\frac{100}{g_*}\right)^{1/4}\sqrt{\mathcal{H}_{wf}~M_P}\sim70$ GeV, using Eq.(\ref{eq:Trh}). 

\medskip

\section{Weakly Coupled $B-L$ portal Dark Matter}



Due to the choice of the tiny gauge coupling $g_{BL}$, the DM particle has never been in thermal equilibrium.
Hence we consider the freeze-in mechanism for producing DM particle via the $Z_{BL}$-portal. 
We will provide a benchmark point which satisfies all the inflationary, reheating \& DM relic density constraints and will also discuss a with possibility to hunt for the $Z_{BL}$ gauge boson in future experimental facilities.




The dark matter physics in our scenario is very similar to the case considered in Ref.\cite{Mohapatra:2019ysk}, and thus we follow the freeze-in DM scenario 
along with the conditions that the $Z_{BL}$ was in thermal equilibrium with SM particles with $g_{BL}\geq 2.7 \times 10^{-8} \sqrt{m_\chi[{\rm GeV}]}$ (see 
Ref.\cite{Mohapatra:2019ysk} for details).
We assume zero initial abundance of DM at the time of reheating after inflation. There are two processes responsible for the DM production,
  $ f \bar{f} \to \chi \bar{\chi} $ mediated by $Z_{BL}$ and $ Z_{BL} Z_{BL} \to \chi \bar{\chi}$, where $f$ denotes SM fermions, and the corresponding cross sections are given by
\begin{eqnarray} 
&& \sigma (\bar{\chi} \chi \to f \bar{f} )\, v \simeq  \frac{37}{36 \pi s}  (Q_{\chi}  g_{BL})^2 g_{BL}^2,  \nonumber\\
&& \sigma (\bar{\chi} \chi \to Z_{BL} Z_{BL})\, v \simeq  \frac{(Q_{\chi}  g_{BL})^4}{4 \pi s } \left( \ln \left[\frac{s}{m_\chi^2} \right] -1 \right), 
\label{DMSigma}
\end{eqnarray} 
assuming $m_b^2 \ll m_\chi^2 < m_t^2$ and $m_{Z_{BL}}^2 \ll m_\chi^2$, $m_b$ and $m_t$ being the masses of the bottom and the top quarks respectively. Here, we have used the approximation formulae in Eq.~(\ref{DMSigma}) as it was shown in Ref.\cite{Mohapatra:2019ysk} to produce almost the same results as numerical computations. 

Using these scattering processes, the DM relic ($\Omega_{CDM} h^2 = 0.12$) constraint can be translated to the following relations (for details, see Ref.\cite{Mohapatra:2019ysk}):
\begin{eqnarray}
  (Q_{\chi}  g_{BL})^2 \, g^2_{BL} + \frac{0.82}{1.2} \, (Q_{\chi}  g_{BL})^4 \simeq 8.2 \times 10^{-24}. 
 \label{caseA}
\end{eqnarray}  
which is insensitive to the ${Z_{BL}}$ boson mass. One can satisfy the DM relic density constraint by
suitably choosing $Q_{\chi}$ (see Fig.\ref{fig:collider}).  For our choice of $g_{BL}=10^{-4}$, Eqn. \ref{caseA} is satisfied with $Q_{\chi}=3\times10^{-4}$.

\section{Experimental Probes}

\begin{figure}
\begin{center}
\includegraphics[height=0.55\textwidth]{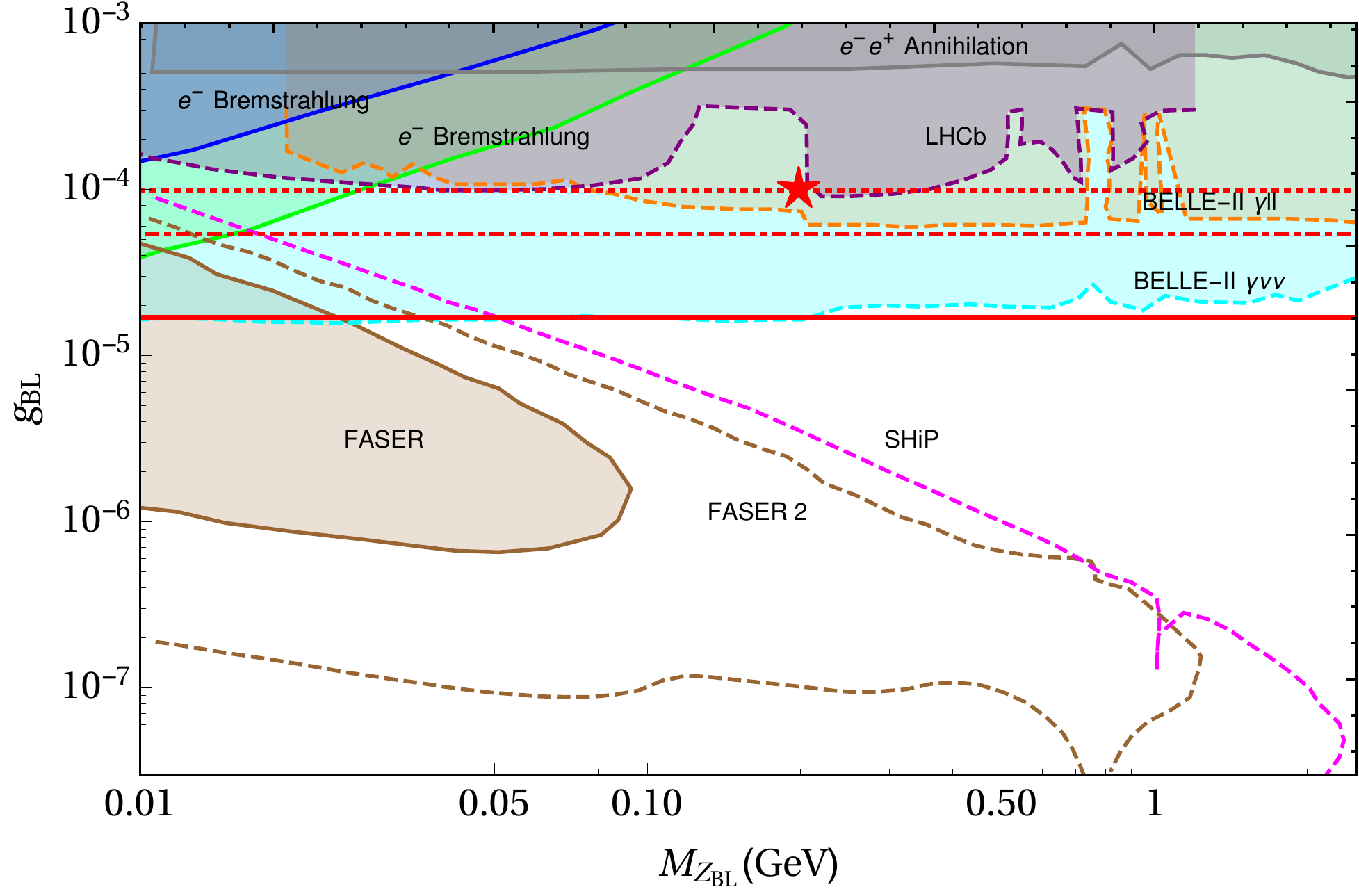}
	\caption{\it  Plot of parameter space of $Z_{BL}$ searches. The three horizontal red lines from bottom most to top most represent $\Omega_{DM} h^2= 0.12$ contours where $Q_{\chi}$ values are respectively $10^{-2}$ (solid), $10^{-3}$ (dotdashed) and $3 \times 10^{-4}$ (dotted). Electron breammstrahlang and annihilation results are obtained from $\rm {DarkCast}$ \cite{Ilten:2018crw,Deppisch:2019kvs}. We show the sensitivity reaches of the several experiments in different color lines. FASER  and FASER 2 \cite{Ariga:2018uku} in solid and dashed brown lines, ShiP \cite{Alekhin:2015byh} in magenta dashed line, light-blue and orange dashed lines for Belle II \cite{Dolan:2017osp}, and purple dashed line for LHCb \cite{Ilten:2015hya,Ilten:2016tkc}. The point marked with red \textbf{$\star$} represents the benchmark point we chose, as an example, where inflationary, reheating and DM relic density constraints are satisfied. 
	}
\label{fig:collider}
\end{center}
\end{figure}

In order to understand a possible pathway to probe our 
scenario, we consider the various laboratory-based experiments, which are the so-called lifetime frontier experiment to search for long-lived particles ($Z_{BL}$ gauge boson in our model).
 These searches are based on the possibility that a new particle produced at the colliders travels and decays, exhibiting displaced vertex and/or missing energy signatures. 
We presented in Fig.\ref{fig:collider} the current experimental constraints and future sensitivity-reaches in the ($M_{Z_{BL}}, g_{BL}$) - plane. Here, the experiments we consider are FASER and FASER 2 \cite{Ariga:2018uku}, SHiP \cite{Alekhin:2015byh}, Belle II and LHCb \cite{Dolan:2017osp,Ilten:2015hya,Ilten:2016tkc}. The planned FASER detector
\footnote{FASER detector is in a tunnel near the ATLAS detector of the LHC, and about 480 meters away to look for displaced vertices with charged particles from arising from long-lived neutral particles produced at the primary vertex of the LHC.} and SHiP \cite{Alekhin:2015byh,Ariga:2018uku} at the LHC will be able to probe the low $M_{Z_{BL}}$ \& low $g_{BL}$ values.
In Fig. \ref{fig:collider}, the horizontal red lines (solid, dot-dashed \& dotted) correspond to the results for the various $Q_{\chi}$ values of the DM particle $\chi$, satisfying $\Omega_{CDM} h^2= 0.12$. For higher $g_{BL}$ \& low $M_{Z_{BL}}$ values, the
electron breammstrahlang modes provide tighter constraints, whereas for higher
$g_{BL}$ \& $M_{Z_{BL}}$ values, the electron-positron annihilation mode is more stringent (for details see Ref.\cite{Ilten:2018crw,Deppisch:2019kvs}). Our benchmark point, denoted by a red $\star$ in Fig.\ref{fig:collider}, will be within the reach of BELLE-II searches. 


\section{Discussions \& Conclusions}
\label{Conclusions}

We investigated a minimal $B-L$ extension of the SM model, and predicted the parameter space where the problems of dark matter, neutrino masses, inflation \& reheating can be accounted for. We summarize our findings in the following points below:
\begin{itemize}
    \item A very low-scale inflation can be driven by the effective flat potential of the $B-L$ Higgs field $\phi$ (which breaks  due to bosonic \& fermionic quantum corrections, in which the inflaton is extremely light: $m_\phi\sim6\times10^{-9}$ GeV for our benchmark point. This raises an issue that the reheating temperature from the inflaton decay is too low to satisfy the model-independent lower bound on $T_{re} \ge 1$ from the successful Big Bang Nucleosynthesis. We showed that in \textit{2-field} system, due to the elliptic shape of the valley near the minima, the
     field oscillates also in the SM Higgs direction (see Fig. \ref{fig:phi and h}), and
     the resulting effective decay rate is much
     larger than the inflato decay rate.
     \item Analysing the field dynamics after inflation near the minima of the \textit{2-field} system, we provided an approximate formula for the reheating temperature,
     given by (Eq. \ref{eq:Trh}), which is entirely 
     different from the standard inflaton decay rate.
     We emphasize the fact that the approximate formula presented in this paper can be used for generic inflation models which involve 2-fields in the oscillation epoch after inflation and a naive estimate of the reheating temperature from the inflaton decay is extremely low. Using this \textit{2-field} analysis as the re-heating mechanism the actual reheat temperature (in the standard inflaton decay case) if found to be much higher than the one naively estimated by the original inflaton decay width. For our inflationary benchmark point the reheating temperature is estimated to be $T_{rh} \sim 70$ GeV, even in the case of negligible decay rate of the inflaton itself.
    \item We investigated the $Z_{BL}-portal$ DM in a minimal U(1)$_{B-L}$ extension of the SM, where the $B-L$ gauge coupling is determined by the requirement from the inflationary observables.For our benchmark parameter choice , the observed DM relic density is reproduced.
    \item We considered the search for the Z$_{BL}$ gauge boson in current \& future lifetime frontier experiments, and speculated that the parameter space that satisfies inflationary, reheating and dark matter constraints simultaneously, can be explored in the future. In Fig.\ref{fig:collider}, we give one bench-mark
    point of $m_{Z_{BL}} = 200$ MeV \& $g_{BL} = 10^{-4}$ which will be within the reach of next generation collider experiments.
\end{itemize}

In future, we will look to build upon our studies, and understand other SM extensions as a direction of model-building for inflation, reheating, dark matter and neutrino mass frameworks especially involving light dark sector. The analytical formula for the re-heating temperature we obtained in our studies can be used in several light inflaton models which offer several interesting and particularly complementary experimental probes in the context of cosmological, astrophysical and laboratory based searches for new physics beyond the SM. 





\section{Acknowledgements}


This work is supported in part by the United States, Department of Energy Grant No. DE-SC0012447 (N.O.). 


\bibliographystyle{apsrev4-1}
\bibliography{ref}
\end{document}